\documentstyle[dina4,epsf,twocolumn]{article}
\begin{document}
\renewcommand{\textfraction}{0.0}
\renewcommand{\topfraction}{1.0}
\renewcommand{\bottomfraction}{1.0}
\def\figurename{Fig.}
\rule[-8mm]{0mm}{8mm}
\begin{minipage}[t]{16cm}
{\large \bf MAGNETIC PROPERTIES OF THE 2D  
$ \bf t$--$\bf t'$--HUBBARD  MODEL\\[4mm]}
U. Trapper$^1$,  H.~Fehske$^2$ and D. Ihle$^1$\\[3mm]
$^1$Institut f\"ur Theoretische Physik, Universit\"at Leipzig, D--04109
Leipzig, Germany\\
$^2$Physikalisches Institut, Universit\"at Bayreuth, D--95440 Bayreuth,
Germany\\[4.5mm]
\hspace*{0.5cm}
  The two--dimensional (2D) t--t'--Hubbard model is studied within
  the slave--boson (SB) theory. At half--filling, a paramagnetic to
  antiferromagnetic phase transition of
  first order at a finite critical interaction strength $U_c(t'/t)$ is
  found. The dependences on $U/t$ and $t'/t$ of the sublattice
  magnetization and of the local magnetic moment are calculated. 
  Our results reasonably agree with recent (Projector) Quantum Monte 
  Carlo data. The SB ground--state phase diagram reveals a 
  $t'$--induced electron--hole asymmetry, and, depending on the ratio 
  $t^\prime /t$, the antiferromagnetic or 
  ferromagnetic phases are stable down to $U=0$ at a critical hole doping. 
\end{minipage}\\[4.5mm]
\normalsize
The magnetic behaviour of strongly correlated itinerant electron
systems, in particular of high--$T_c$ cuprates, is frequently
described on the basis of the one--band Hubbard model with nearest
$(t)$ and next--nearest neighbour 
hopping $(t^\prime)$~[1-5]. In this work we explore
the ground--state properties of the 2D t--t'--Hubbard model
which can be expressed in the four--field SB representation~[6] as
\begin{equation}
{\cal H}= \sum_{ ij \sigma}
t_{ij}^{ }
z_{i\sigma}^{\dagger}z_{j\sigma}^{ } 
f_{i\sigma}^{\dagger}f_{j\sigma}^{ } + U\,\sum_i d_{i}^{\dagger}d_{i}^{ } \,.
\end{equation}
Neglecting charge--density--wave states, in the two--sublattice 
(AB) saddle--point approximation, the
free energy per site is given by
\begin{eqnarray}
 f(n,T)\!\!\! &=&\!\!\! U d_{}^{2} 
-\sum_{\eta=\pm}\lambda_{\eta}^{(2)}
(p_{\eta}^{2}+d_{}^{2}) + \mu n  \nonumber \\
&& \hspace*{-0.5cm}+\frac{2}{\beta N}\sum_{\vec{k}\nu=\pm} 
\ln\left[1-f(E_{\vec{k}\nu}^{ }-\mu)\right] 
\end{eqnarray}
with the quasiparticle tight--binding bands
\begin{eqnarray}
E_{\vec{k}\nu}^{ }\!\!\!\! &=&\!\!
\mbox{$\frac{1}{2}$}\sum_{\eta} 
\left[\lambda_{\eta}^{(2)}
+ \mbox{$\frac{1}{2}$}q_{\eta}^{ }
(\varepsilon_{\vec{k}}+\varepsilon_{\vec{k}-(\pi,\pi)})\right] \nonumber \\
 & &\!\!\!\!+\nu \sqrt{A_{}^{2}
+\mbox{$\frac{1}{4}$}q_{+}^{ }q_{-}^{ }
(\varepsilon_{\vec{k}}-\varepsilon_{\vec{k}-(\pi,\pi)})^{2}}\,,
\end{eqnarray}
where the $\vec{k}$--sum runs over the magnetic Brillouin zone, 
the $\lambda_{\eta}^{(2)}$ ensure the SB constraints,
$A= \mbox{$\frac{1}{2}$}\sum_{\eta}\eta [
\lambda_{\eta}^{(2)}
+ \mbox{$\frac{1}{2}$}q_{\eta}^{ }
(\varepsilon_{\vec{k}}+\varepsilon_{\vec{k}-(\pi,\pi)})]$,
$q_{\eta}=|z_{\eta}|^2$, and
$\varepsilon_{\vec{k}}=-2t(\cos k_{x}^{ }+ \cos k_{y}^{ })
-4 t' \cos k_{x}^{ } \cos k_{y}^{ }\,$.

At half--filling and $t^\prime\neq 0$, we obtain a 
paramagnetic (PM)$\rightleftharpoons$antiferromagnetic (AFM) phase
\begin{figure}
\rule[-5.8cm]{0mm}{6.36cm}
\end{figure}
transition of first order at the critical interaction strength
\begin{figure}[b]
\centerline{\mbox{\epsfxsize 7.2cm\epsffile{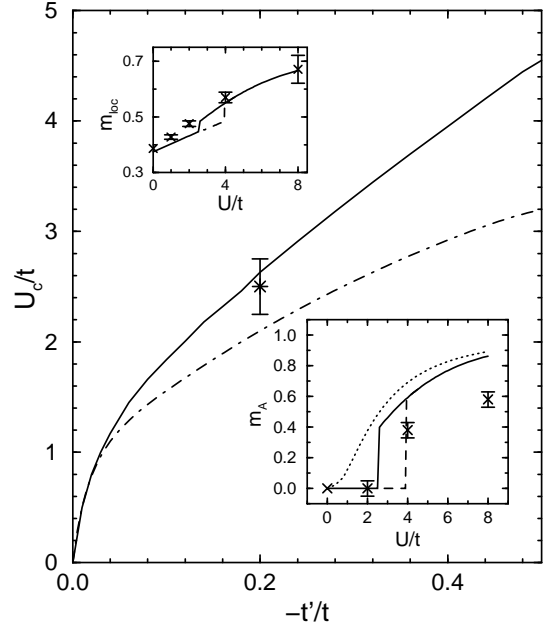}}}
\caption{$U_c$ {\it vs} $t^\prime$ for the PM to AFM transition 
compared with HF (chain dashed) and QMC ($\ast$) 
results. The insets show $m_A$ and $m_{loc}$ {\it vs} $U$ 
at $t^\prime /t=-0.2$ (solid) and -0.4 (dashed) together with the 
Projector QMC ($\times$) and spin--wave data (dotted) at $t^\prime
/t=-0.2$ taken from Ref.~[4].}
\end{figure}
$U_c(t^\prime /t)$ (see Fig.~1) which is accompanied by the opening of
the indirect gap in the SB band structure~(3) shown in Fig.~2.
At $t^\prime /t=-0.2$ we get $U_c/t=2.63$ which reproduces the Quantum
Monte Carlo (QMC) value~[1]. This result may be explained by the shift
of the logarithmic van Hove singularity for $t^\prime \neq 0$.
\begin{figure}[t]
\begin{minipage}[b]{16cm}
\centerline{\hspace*{0.5cm}\mbox{\epsfxsize 5cm\epsffile{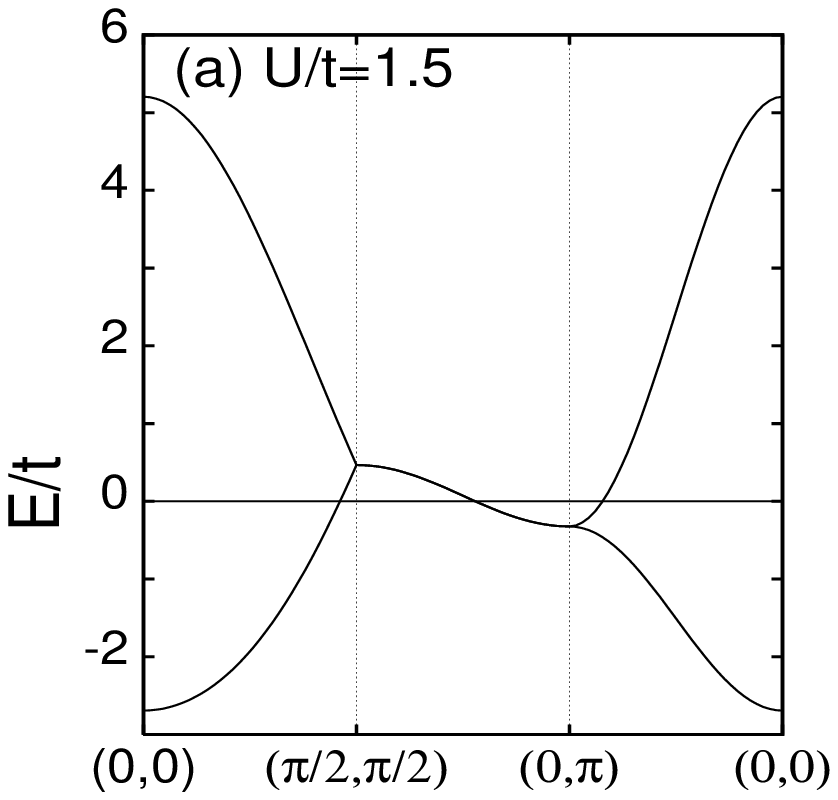}}
            \mbox{\epsfxsize 5cm\epsffile{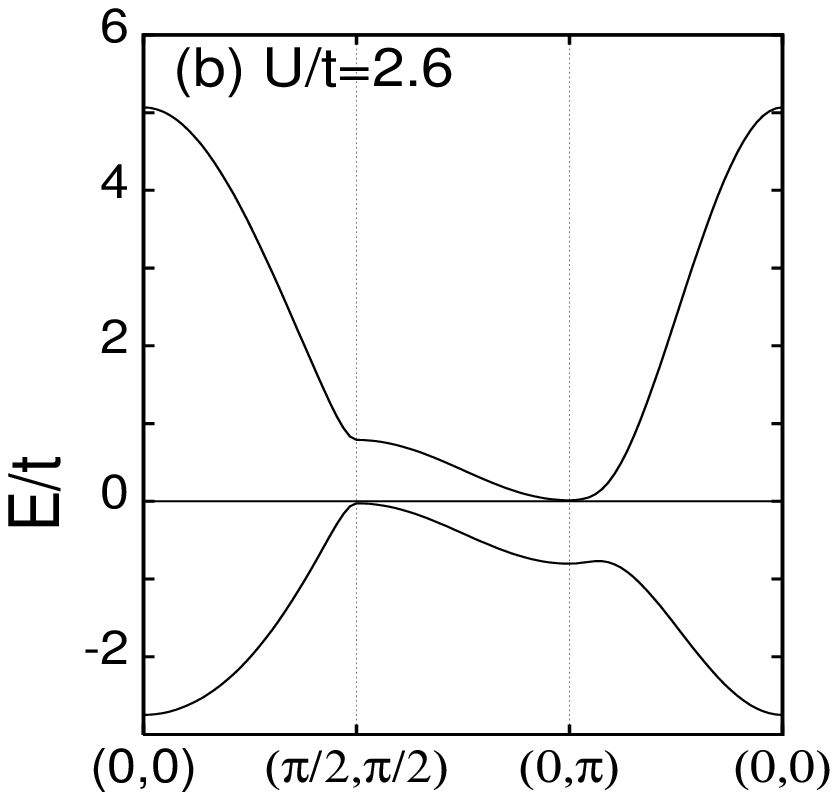}}
            \mbox{\epsfxsize 5cm\epsffile{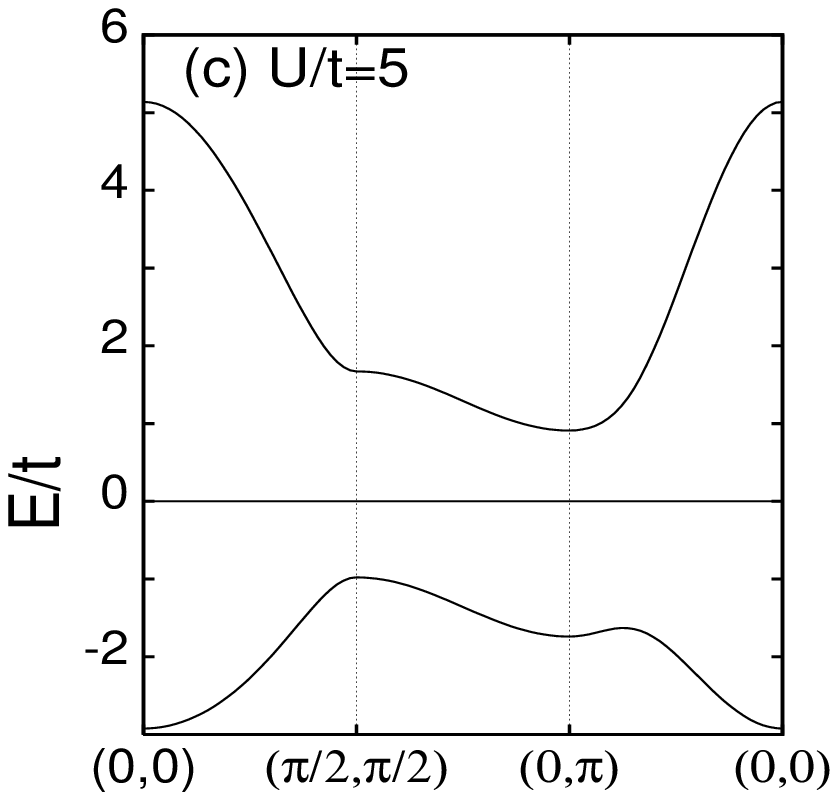}}}
\caption{Band dispersion $E_{\vec{k}\nu}-\mu$ at $n=1$
and $t^\prime /t=-0.2$ for $U<U_c$ (a), $U=U_c$ (b), and $U>U_c$ (c).}
\end{minipage}\\[-0.5cm]
\end{figure}
Note that in our SB calculation a metallic AFM ground
state, as suggested recently in Ref.~[3], does not exist. Of course,
such a phase may be stabilized introducing additional hopping
terms by hand~[3].
\begin{figure}[b]
\centerline{\mbox{\epsfxsize 7.5cm\epsffile{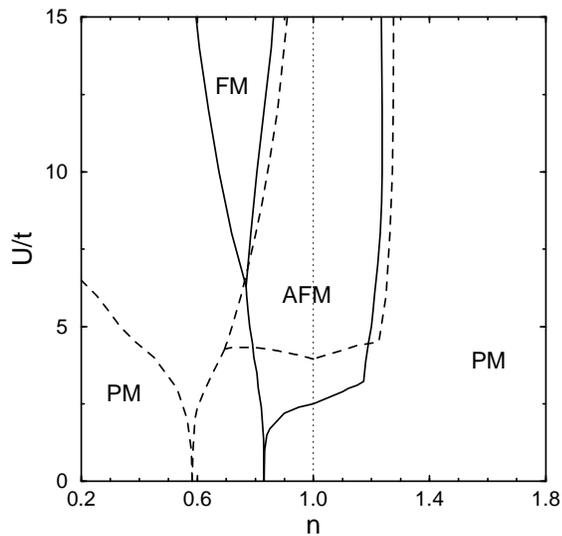}}}
\caption{Phase diagram of the t--t'--Hubbard model at
$t^\prime /t=-0.2$ (solid) and $t^\prime /t=-0.4$ (dashed).}
\end{figure}
As seen in Fig.~1, the sublattice magnetization
$m_A^{}=p^2_{+}-p^2_{-}$ and the local magnetic
moment $m_{loc}=\frac{3}{4}(n-2d^2)$ reasonably agree
with Projector QMC data at $t^\prime /t =-0.2$~[4]. 

The SB
ground--state phase diagram depicted in Fig.~3 reveals a pronounced
\mbox{$t^\prime$--induced} electron--hole asymmetry and the stability of the
AFM state  (at $t^\prime /t=-0.2$) and
of the ferromagnetic (FM) state 
(at $t^\prime /t=-0.4$) down to $U=0$ at the critical hole
dopings 0.17 and 0.418, respectively. 
This qualitatively agrees with the Hartree--Fock (HF) calculation
of Ref.~[1], but contradicts the HF solution obtained in Ref.~[5].
\begin{figure}[t]
\rule[-6.05cm]{0mm}{2.5cm}
\end{figure}
Compared with the HF results, the electron correlations incorporated
in the SB approach reduce the stability regions of the long--range
ordered phases in the favour of the PM phase.

In conclusion, the magnetic ground--state properties of the
t--t'--Hubbard model are well described by 
our SB approach, in particular the weak--interaction limit is correctly
reproduced.

This work was supported by the DFG under project
SF--HTSL--SRO. U.T acknowledges the hospitality at the University Bayreuth.
We thank M. Deeg for help with the computer work. 
\\[0.4cm]
REFERENCES
\begin{enumerate}
\item H.~Q. Lin and J.~E. Hirsch, Phys. Rev. B {\bf 35}, 3359 (1987).
\vspace*{-0.2cm}
\item U.~Trapper, D.~Ihle, and H.~Fehske, Phys. Rev. B {\bf 52}, R 11553 
(1995); Int. J. Mod. Phys. B, accepted for publication (1996).
\vspace*{-0.2cm}
\item D. Duffy and A. Moreo, cond-mat/9604172.
 \vspace*{-0.5cm}
\item R.~E. Hetzel, P.~Topalis, and K.~W. Becker, Physica B, to be
 published (1996).
\vspace*{-0.2cm}
\item W.~Brenig, J. Low Temp. Phys. {\bf 99}, 319 (1995).
\vspace*{-0.2cm}
\item G. Kotliar and A. E. Ruckenstein, Phys. Rev. Lett {\bf 57}, 
1362 (1986)
\vspace*{-0.3cm}

\end{enumerate}
\end{document}